\documentclass[epj]{svjour}
\begin{document}
\title{Low temperature penetration depth and the effect of
quasi-particle scattering measured by millimeter wave transmission
in YBa$\bf _2$Cu$\bf _3$O$\bf _{7-\delta}$ thin films}
\author{S.~Djordjevic, L.A.~de~Vaulchier, N.~Bontemps, J.P.~Vieren,
Y.~Guldner\inst{1},
S.~Moffat, J.~Preston\inst{2},
X.~Castel, M.~Guilloux-Viry \and A.~Perrin\inst{3}
}
%
%
\institute{Laboratoire de Physique de la Mati\`ere Condens\'ee, Ecole Normale
Sup\'erieure, 24 rue Lhomond, 75231 Paris cedex 05 (France) \and
Brockhouse Institute for Materials Research, McMaster University, Hamilton On.
L8S 4M1 (Canada) \and
Laboratoire de Chimie du Solide et Inorganique Mol\'eculaire, Universit\'e de
Rennes~I, Av. du G\'en\'eral Leclerc, 35042 Rennes cedex (France)}
\titlerunning{Low temperature penetration depth and the effect of
quasi-particle scattering...}
\authorrunning{S.~Djordjevic {\it et al.}}
\date{Received: date / Revised version: date}
%
\abstract{
Measurement of the penetration depth $\lambda(T)$ as a function of temperature
using millimeter wave transmission in the range 130-500$\,$GHz are reported for
three $\rm YBa_2Cu_3O_{7-\delta}$ (YBCO) laser ablated thin films. Two films,
deposited on a $\rm LaAlO_3$ substrate ($T_c = 90.2\,\rm K$), exhibit a narrow
resistive transition ($0.3\,\rm K$). One has been subsequently irradiated with
$\rm He^+$ ions in order to increase the scattering rate of the quasi-particles
($T_c = 87.8\,\rm K$). The third film, grown on MgO ($T_c = 88.5\,\rm K$),
exhibits also a fairly narrow transition ($0.8\,\rm K$) and a high crystalline
quality. The experiment provides the absolute value $\lambda(T \leq 30\,\rm K)$
for the penetration depth at low temperature: the derivation from the
transmission data and the experimental uncertainty are discussed. We find a zero
temperature penetration depth $\tilde{\lambda_0} = 1990 \pm 200\,\rm \AA$, 
$2180 \pm 200\,\rm \AA$ and $2180 \pm 200\,\rm \AA$, for
YBCO-500\AA /$\rm LaAlO_3$ (pristine), YBCO-1300\AA /MgO and YBCO-500\AA
/$\rm LaAlO_3$ (irradiated) respectively. $\lambda(T \leq 30\,\rm K)$ exhibits a
different behavior for the three films. In the pristine sample,
$\lambda(T \leq 30\,\rm K)$ shows a clear {\it temperature and frequency
dependence}, namely the temperature dependence is consistent with a linear
variation, whose slope decreases with frequency: this is considered as an
evidence for the scattering rate being of the order of the measuring frequency.
A two fluids analysis yields
$1/\tau(T \leq 30\,\rm K) \sim 1.7 \times 10^{12}\,s^{-1}$.
In the two other samples, $\lambda(T \leq 30\,\rm K)$ does not display any
frequency dependence, suggesting a significantly larger scattering rate. The
temperature dependence is different in these latter samples. It is consistent
with a linear variation for the YBCO/MgO sample, not for the YBCO/$\rm LaAlO_3$
irradiated one, which exhibits a $T^2$ dependence up to $40\,\rm K$. We have
compared our data to the predictions of the $d$-wave model incorporating
resonant scattering and we do not find a satisfactory agreement. However, the
large value of $\tilde{\lambda_0}$ in the pristine sample is a puzzle and sheds
some doubt on a straightforward comparison with the theory of data from thin
films, considered as dirty $d$-wave superconductors.
} 
\maketitle
\section{Introduction}
\label{introduction}
The study of the in-plane penetration depth $\lambda(T)$ and its temperature
dependence has been widely investigated since the discovery of high-$T_c$
cuprates. Being directly related to the superfluid density, it probes the
density of excited quasi-particles (QP), and therefore it is closely related to
the symmetry of the gap. One of the key experiment in favor of an
unconventional order parameter with lines of nodes was the evidence of
a linear temperature dependence of $\lambda(T)$ in YBCO single crystals
\cite{hardy93}. Such a linear dependence has been now observed in a variety of
cuprates, e.g. $\rm Bi_2Sr_2CaCu_2O_{8+\delta}$, $\rm Tl_2Ba_2CuO_{6+\delta}$
and $\rm Hg_2Ba_2Ca_{n-1}Cu_nO_{2n+2+2\delta}$,
\cite{jacobs95,lee96,broun97,panagopoulos96}
with slopes ranging from 4 to $13\,\rm \AA K^{-1}$. This behavior is consistent
with a $d_{x^2-y^2}$ order parameter which is presently strongly in favor,
supported by photoemission \cite{ding95} and phase sensitive data in
$\rm Bi_2Sr_2CaCu_2O_{8+\delta}$, YBCO,  $\rm Tl_2Ba_2CuO_{6+\delta}$
\cite{koutnetzov97}. However, the precise symmetry of the order
parameter is still controversial or may differ in the various cuprate
families, in particular in YBCO, $\rm Bi_2Sr_2CaCu_2O_{8+\delta}$
or $\rm La_{2-x}Sr_xCu_{4+\delta}$ \cite{maki95,bouvier95,achaf97}.
In YBCO, suggestions have been brought up from experiments, including
the admixture of a significant $s$-wave component \cite{koutnetzov97} and
the claim for a two-components order parameter \cite{srikanth97}. Going
beyond qualitative tests of the symmetry is difficult. It requires a
model providing quantitative predictions. Such models have been developed.
We mention here some of them, namely pure $d_{x^2-y^2}$ order
parameter in the framework of the spin fluctuations theory \cite{monthoux93},
$s+d$ symmetry related to YBCO orthorhombic structure \cite{maki95},
anisotropic $s$-wave \cite{bouvier95} or {\it plane-chain coupling} in
YBCO \cite{combescot95}.
There is a large body of literature which discusses in detail the
microwave behavior of a $d$-wave superconductor in the presence of
scattering. We briefly recall those features of the model which deal
with the penetration depth and the high frequency conductivity.
The change of i) the critical temperature $T_c$, ii) the superfluid
density at zero temperature $\rho_s \propto 1/\tilde{\lambda_0^2}$ where 
$\tilde{\lambda_0}$ is the zero temperature penetration depth, 
iii) the temperature dependence of  $\lambda(T)$, has been
calculated as a function of the impurity concentration and/or the
scattering rate in the unitary limit
\cite{kim94,hirschfeld93,hirschfeld93b,hirschfeld94}. This latter limit may
explain why a very small amount of
impurities or defects responsible for elastic scattering, while hardly
affecting the critical temperature, strongly modifies the superfluid
density, hence the penetration depth and its temperature dependence.
Two regimes are expected to develop successively: the first one
(the so-called gapless regime) holds from low temperature up to a
crossover temperature $T^*$, and the temperature dependence of
$\lambda(T)$ is $T^2$. The second one is the clean regime
where the linear temperature dependence is recovered
\cite{hirschfeld93,hirschfeld93b,hirschfeld94}.

Presently, it seems clearly settled that the low temperature dependence
of $\lambda(T)$ in YBCO single crystals ({\it taken henceforward as
examples of the clean system})
is linear up to $\sim 30\,\rm K$ with a slope of 4 to 5
$\rm \AA \, K^{-1}$ \cite{hardy93,bonn95,mao95}. Thin films have been
considered as possible examples of ``dirty'' samples, although there is
evidence for a linear temperature dependence with a slope similar to
that of crystals \cite{devaulchier96,froelich94,lemberger96}.
Deviations from this linear regime is however commonly reported in
thin films, whether the linear dependence does not display the
``canonical'' slope \cite{shibauchi96,hensen97,farber} or the dependence
is rather $T^2$ \cite{annett91,ma93,lee94,porteanu95,feenstra97}.
The latter behavior has
been argued to arise from defects likely to be present in thin films
and playing the role of impurities in the unitary limit. A quantitative
comparison to the $d$-wave model was claimed to apply successfully to such
samples \cite{annett91,ma93}. A similar comparison has been performed
in purposely impurity-doped YBCO single crystals \cite{bonn94}. Note
that the experimental techniques used in most of the above mentioned
reports provide the change of $\lambda(T)$ with temperature, but
not its absolute value with a sufficient accuracy. The single crystal
data have been analyzed by Hirschfeld {\it et al.} within the framework
of the $d$-wave model. The conclusion was that, as long as the absolute
value $\tilde{\lambda_0}$ was not known, ``roughly equally good fits'' may
be obtained
with various scattering rates, the latter being consistent or not
with the initial assumption of the scattering rate scaling with the
impurity concentration. Some of the thin film data did not display
a satisfactory agreement with $d$-wave theory either \cite{ulm95}.
Similar restrictions were also raised by Hensen {\it et al.} when
analyzing their own thin film data \cite{hensen97}. 

We report in this paper penetration depth results obtained in three
selected thin films (which we describe specifically in Section~\ref{samples}).
The penetration depth is deduced from far infra red (FIR) transmission
of an electromagnetic wave through the sample, one technique among very
few others which yields not only the variation of $\lambda(T)$
with temperature, but its zero temperature value $\tilde{\lambda_0}$
\cite{devaulchier96,devaulchier95,budhani91}. As already stated, this is
a key parameter for a reliable comparison with theoretical models.
Moreover, the experiment being performed between 130 and $510\,\rm GHz$, at
several frequencies (usually 5), the corresponding time scale becomes
comparable to the  low temperature QP scattering time $\tau$ 
(typically, $300\,\rm GHz$ corresponds to
$1/\tau \sim 2 \times 10^{12}\,\rm s^{-1}$).
This manifests itself through a dependence of the measured penetration
depth as a function of frequency and can be analyzed quantitatively in
the frame work of a generalized two fluids model including a Drude
conductivity for the normal fluid
\cite{shibauchi96,bonn94,bonn93,bonn93b,dahne95}. We demonstrate that our data
display a satisfactory agreement with this phenomenology and extract the low
temperature ($T \le 30\,\rm K$)
elastic scattering rate. This is in contrast with earlier work where
the internal consistency of the two fluids model could not be checked
since the measurements were performed as a function of temperature, at
one or two frequencies. Therefore a two fluids conductivity had to be
assumed in order to deduce the scattering rate either from surface
resistance data in single crystals \cite{bonn93b} or from FIR or THz
spectroscopy in presumably good quality films
\cite{dahne95,nuss91,frenkel96}. Since our experiment yields
$\tilde{\lambda_0}$  and
$\lambda(T)$, we derive from this set of data three parameters of
the model, namely $\delta \lambda_0 = \tilde{\lambda_0} - \lambda_0$
($\lambda_0$ is the penetration depth of the clean system),
$T^*$ and $T_c / T_{c0}$ ($T_c$ and $T_{c0}$
are the actual transition temperature and the one of the clean system
respectively). We shall take  $T_c$ at zero resistance (Table~\ref{table1}) and
$T_{c0} = 92\,\rm K$. We then compare to the values these parameters
should achieve in the case of the $d$-wave model in presence of resonant
scattering. As will be shown, the model clearly fails to account for
the data in all the films under investigation in this paper. Since more
experimental parameters are available from our experiment than from the
others, our results shed some doubt on the previous claims that thin
films can be properly described in the framework of the $d$-wave model
by simply incorporating resonant scattering. 

Our paper proceeds as follows:
In Section~\ref{experimental}, we describe the experimental set-up.
Section~\ref{data} describes
in detail the data processing from which the absolute value of the
penetration depth $\tilde{\lambda_0}$ and its temperature variation
are derived, and introduces the frequency dependent penetration depth
$\lambda(\omega,T)$. The samples are described in Section~\ref{samples}.
Section~\ref{penetration} summarizes the penetration depth data.
In Section~\ref{analysis} we
recall the so-called generalized two fluids model, and stress the
consequences on the variation of $\lambda(\omega,T)$
when the low temperature QP scattering rate
$1/\tau$ becomes comparable to the measuring frequency
$\omega$. This leads us to an estimate of $1/\tau$.
Section~\ref{discussion} is devoted to a comparison of our experimental data
with the $d$-wave model, assuming that thin films may be examples of
$d$-wave superconductors, however including strong scattering due to
their specific defects.
\section{Experimental}
\label{experimental}
We briefly recall the general features of the experimental millimeter
transmission set-up that has been described in detail elsewhere
\cite{devaulchier96,devaulchier95,devaulchier95b} and point out some late
modifications. The sources are carcinotrons operating
typically at 134, 250, 333, 387 and 510$\,$GHz.
The 510$\,$GHz source can be tuned in a typical 50$\,$GHz range. At 387$\,$GHz,
tuning is possible but difficult.
These sources deliver from 1 to 300$\,$mW and are very stable
($\pm 0.1\,\rm dB$ per hour). A continuously tunable 110-180$\,$GHz backward
wave oscillator (BWO) is also available, which provides an output power of
typically 5$\,$mW and a stability of $\pm 0.3\,$dB.
The radiation is guided through oversized (10$\,$mm diameter) circular
wave guides, which allow us to change the frequency, onto the sample
placed in a helium flow cryostat. The sample is tightly fixed with
silver paste (to avoid leakage) on a brass sample holder
(with a 3$\,$mm or 4$\,$mm iris) and is surrounded by a helium exchange gas
(100$\,$mbar pressure at room temperature) to damp out temperature
instabilities in the helium flow at our lowest temperatures. The
transmitted signal is detected by a He cooled InSb bolometer with a
60$\,$dB linearity range.
We have checked that the microwave leakage represents more than 60$\,$dB
attenuation. Samples are screened from stray magnetic field, so that
the residual magnetic field is less than 0.5$\,$Oe.
The main limitation to the accuracy of this experiment is associated
with the multi-mode propagation within the oversized guides, giving rise
to a set of standing waves which are subject to thermal drifts (this
problem, discussed in some detail below, has been carefully analyzed
using the tunable BWO). To minimize those thermal drifts, the guides
within the cryostat are made out of invar, and are as short as possible,
compatible with a good temperature stability of the sample holder
($\pm 0.2 \,\rm K$). The measurements are performed by slowly varying the
temperature from 6$\,$K to 110$\,$K, at fixed frequency.
\section{Data processing}
\label{data}
\subsection{Transmission of a thin film}
\label{transmission}
For the sake of emphasizing the physics of the measurement, we firstly
neglect in this paragraph the interference pattern due to the substrate.
We consider a film of thickness $d$ (smaller than the wavelength of the
electromagnetic wave, the skin depth in the normal state or the
penetration depth in the superconducting state) deposited on a substrate
of index $n$. The transmission reads \cite{glover57}:

\begin{eqnarray}
{T_r(\omega,T)} = {1\over \left| 1+{\sigma(\omega,T) d Z_0 \over 1+n} \right|^2}
\label{eq1}
\end{eqnarray}

where $Z_0 = 377 \,\rm \Omega$ is the impedance of the free space.
$\sigma(\omega,T) = \sigma_1(\omega,T) -i\, \sigma_2(\omega,T)$
is the complex conductivity of the film, and
$Z(\omega,T)=1/\sigma(\omega,T) d$  is the impedance of the film. 
The imaginary part of the conductivity writes:

\begin{eqnarray}
{\sigma_2(\omega,T)} = {1 \over {\mu_0 \omega \lambda_L^2(T)}}
\label{eq2}
\end{eqnarray}
 	
where $\lambda_L(T)$ is the London penetration depth \cite{tinkham75}.
The usual condition for this expression of $\sigma_2(\omega,T)$ 
to be valid is that the working frequency must lie far below the
superconducting (isotropic) gap $\Delta$ ($\omega \ll 2 \Delta$)
\cite{tinkham75}. In cuprates in general, and YBCO in particular,
experiments including the penetration depth data show that the gap
exhibits nodes in some directions of the $k$ space. Therefore
the condition  $\omega \ll 2 \Delta(k)$ is not fulfilled in all
directions of the $k$ space. Nevertheless, the density of states at finite
energy $\omega$ is expected to be small compared to that at the
maximum gap energy $\Delta_m$ \cite{preosti94}, as long as
$\omega \ll 2 \Delta_m$. In this case, we consider that the
electromagnetic response is dominated by the one of the superfluid, and
that the contribution from QP generated through pair-breaking by the EM
wave is negligible. Our highest frequency ($17\,\rm cm^{-1}$) is now to be
compared to $2 \Delta_m$. Tunneling data locate a maximum around 25$\,$meV
\cite{beasley91}. If this feature is assigned to the maximum gap, then
indeed $\omega \ll 2 \Delta_m = 400\,\rm cm^{-1}$.\\
In order now to deduce  $\lambda_L(T)$ from (\ref{eq1}), another condition is
that $\sigma_2(\omega,T)$, as given above, be much larger than
$\sigma_1(\omega,T)$, which is well known at low temperature for
a BCS-type superconductor. In an unconventional superconductor with
nodes in the gap, the conductivity may involve a contribution arising
from thermally excited QP \cite{hirschfeld93b,hirschfeld94,hensen97}. The real
part of the conductivity for a $d$-wave superconductor, in the presence of
inelastic as well as elastic scattering, exhibits a Drude-like expression,
however including now a frequency-dependent QP scattering rate
\cite{hirschfeld93b,hirschfeld94,hensen97}. As further discussed below, these
approaches have not proven completely successful in order to account
for the experimental data. In the following, we shall use two different
frameworks for data analysis: i) as many other authors, a generalized
two fluids conductivity, where the QP scattering rate does not depend
upon the frequency, in order to account for the presence of unpaired
carriers; ii) the $d$-wave model for the conductivity in presence of
elastic scattering due to impurities or defects
\cite{kim94,hirschfeld93,hirschfeld93b,hirschfeld94}.\\
A this stage, we define the conductivity, by including a real part
$\sigma_1(\omega,T)$ arising from the normal fluid and an
imaginary part $\sigma_2(\omega,T)$ 
where the contribution of the normal fluid is included formally in
a frequency and temperature dependent penetration depth
$\lambda(\omega,T)$:

\begin{eqnarray}
{\sigma(\omega,T)} = \sigma_1(\omega,T) - {1\over {\mu_0 \omega
\lambda^2(\omega,T)}}
\label{eq3}
\end{eqnarray}

The dependence of the generalized penetration depth $\lambda(\omega,T)$
on the frequency depends, as will be shown further, on the relative
value of the frequency and the QP scattering rate $1/\tau$. In the quasi-static
regime, e.g. at very low frequency ($\omega \tau \ll 1$), the generalized
penetration depth does not depend on the frequency. In this limit
$\lambda(\omega,T)$ may be identified to the London penetration depth
$\lambda_L(T)$ \cite{bonn95,devaulchier95,bontemps96}.
The conductivity defined in (\ref{eq3}) is then plugged in the transmission
(\ref{eq1}), and the absolute value of the penetration depth and its
temperature dependence may then be deduced as explained in paragraphs
3.2 to 3.4. Note that we have checked the temperature independence of
the index $n$ of the substrate in our frequency range \cite{devaulchier95}.\\
\subsection{Normalized transmission and the basics of penetration depth
            measurements}
\label{normalized}
Equation~(\ref{eq1}) involving the complex conductivity depends upon two
quantities, $\sigma_1(\omega,T)$ and
$\sigma_2(\omega,T)$, and only a single measurement
is available. However, we are interested in the low temperature
behavior of the transmission. In a BCS superconductor, one has
$\sigma_2(\omega,T) \gg \sigma_1(\omega,T)$
(actually even close to $T_c$). In this case, the transmission
depends only on $\sigma_2(\omega,T)$, hence on
$\lambda(\omega,T)$. In high-$T_c$ superconductors, the question
is less simple: it is known from experiments that $\sigma_1(\omega,T)$ 
{\it increases} below $T_c$, exhibits a maximum around 40$\,$K at
frequencies $\leq 50\,\rm GHz$
\cite{shibauchi96,farber,bonn93,bonn93b}, which shifts toward
higher temperature at higher frequency (typically 60$\,$K at 300$\,$GHz
as measured in thin films \cite{hensen97,nuss91,frenkel96}). It is
therefore not obvious that $\sigma_1(\omega,T)$ is
negligible with respect to $\sigma_2(\omega,T)$ in our
frequency range. All data in the literature show however that, at low
temperature, e.g. below 30$\,$K, which is the range of interest in the
present paper, the approximation
$\sigma_1(\omega,T) \ll \sigma_2(\omega,T)$ holds
\cite{bonn95,shibauchi96,hensen97,farber,bonn93,bonn93b,nuss91,frenkel96}
hence $T_r(\omega,T)$ is eventually simply related to the penetration depth.\\
The absolute value of the transmission, $T_r(\omega,T)$ cannot be
determined in our set-up. Such an absolute measurement is possible
in a set-up which allows quasi-optical propagation
\cite{feenstra97,volkov85}. However, in our case, replacing the
sample by a hole or a bare substrate changes the standing waves
within the guides, thus biasing the measurement of the reference
incident power. In order to circumvent the difficulty, we choose
to normalize the transmission with respect to the one in the normal
state at 110$\,$K:

\begin{eqnarray}
{T_r(\omega,T) \over T_r(110\,\rm K)} =
{\Bigl(1+{\sigma(110\,\rm K \it) d Z_0 \over 1+n}\Bigl)^2
\over
1+\Bigl({d Z_0 \over (1+n) \mu_0 \omega \lambda^2(\omega,T)}\Bigl)^2}
\label{eq4}
\end{eqnarray}

which will be henceforward referred to as the normalized transmission
$T_n$. The normalization makes sense (e.g. the normalized transmission
yielding the penetration depth value as shown in (\ref{eq4})) only if the
transmission is independent of the frequency in the normal state at 110$\,$K.
This turns out to be right in our frequency range: considering the value
of the scattering rate $1/\tau \sim 2 \times 10^{13}\,\rm s^{-1}$ in the
normal state (at $T \geq T_c$) inferred from surface impedance data
\cite{bonn94,bonn93,bonn93b}, the condition $\omega\tau \ll 1$ is
verified and the conductivity is frequency independent. Finally, the
choice of 110$\,$K is a compromise for a temperature high enough above
$T_c$ so as to avoid conductivity fluctuations which may be
frequency dependent (but are still negligible at 110$\,$K) and low enough
in order to minimize the thermal drifts due to thermal expansions of
the guides (which are negligible at 110$\,$K).
\subsection{Interference pattern, derivation of $\tilde{\lambda_0}$
            and $\lambda(\omega,T)$}
\label{interference}
(\ref{eq4}) does not account for the interference fringes due to the
substrate. In experiments using quasi-optical techniques, the
interference pattern is very well identified
\cite{feenstra97,dahne95,volkov85}. From the calculation of this
pattern in the case of a plane wave \cite{glover57}, we know that the
contrast of the fringes increases considerably with frequency: it is
10 times larger at 500$\,$GHz than at 130$\,$GHz (Fig.~\ref{fig1}). Therefore,
we expect an interference effect to be more conspicuous at our largest
frequency. Since we have no broadly tunable source in the range 300-500$\,$GHz,
we cannot measure directly this pattern. Moreover, in our set-up, due
to the multi-mode propagation, it is not clear to which extent this
pattern may be modified. We therefore compare the transmission data to
the theoretical transmission including the interference effect, in the
simple case of a plane wave. If the interference effect were negligible,
the data should follow the simplified transmission resulting from (\ref{eq4}).
Fig.~\ref{fig1} displays a set of normalized transmission data from sample MA
for the various frequencies $f = \omega /2\pi$ ranging from
130 to 510$\,$GHz, as a function of $f^2$, at 10$\,$K (although the
penetration depth data will be reported in the range 6$\,$K-110$\,$K, we have
collected many more data starting from 10$\,$K). Clearly, despite the
experimental scatter, one cannot fit the whole set of data with a
straight line (Eq.~\ref{eq4}),
due to the highest frequency data. This strongly suggests that
the interference effect must be taken into account.\\
Let us comment firstly about the experimental scatter. The experimental
points at each frequency correspond to different sets of experimental
runs, e.g. the sample has been removed then put back into the cryostat.
This affects the standing waves by changing  the boundary conditions.
We have established, after a detailed study, that this scatter
is due to thermal drifts occurring above 40$\,$K as
the temperature is swept from 6$\,$K to 110$\,$K. This obviously affects the
normalized transmission since the normalization is performed with respect 
to the 110$\,$K transmission.
Fig.~\ref{fig1} shows the best fit of our data with a given pattern. The
agreement is obtained by adjusting the value of the penetration  depth
and of the thickness $l$ of the substrate. We find:
$\lambda(\omega,10\,\rm K) = 2000 \pm 100\,\rm \AA$ (the frequency dependence
is neglected, this assumption being reasonable at low enough
temperature \cite{hirschfeld93b,hirschfeld94,bontemps96}) and
$l = 515 \pm 10\,\rm\mu m$, the nominal thickness being
$500\,\rm\mu m$. The uncertainty on the penetration depth arises
mainly from the scatter of the experimental points. 
To assert this fit, the value of the thickness $l$ of the substrate
was: i) double checked by a direct measurement
(also within $\pm 10\,\rm \mu m$) with a micrometer screw ii) compared
with the value obtained from the fringe pattern observed in a
transmission measurement performed at LURE using far infrared
synchrotron radiation in the range $20-45\,\rm cm^{-1}$
(yielding the optical path $2nl$),
combined with an infrared measurement of $n$ \cite{lobo,roy}. 

Obviously, we do not use this fitting procedure to establish the value
of $\lambda(\omega,T)$ at each temperature, since its variation
up to 30$\,$K is expected to be small compared to the uncertainty
(taking the ``canonical'' value of $4\,\rm \AA K^{-1}$ from single crystals).
We can however derive with a much better accuracy the relative change
of $\lambda(\omega,T)$ as a function of temperature, starting
from 6$\,$K, since there are no thermal drifts in this temperature range.
We proceed as follows.
Given the best fit at 10$\,$K (Fig.~\ref{fig1}), a factor $\beta_0(f)$ for
each experimental frequency $f$ defines the correspondence between the
actual data (including the interference effect) and the simplified
expression for the normalized transmission neglecting the interference
effect (satisfying (\ref{eq4})). We multiply the temperature dependent
normalized transmission by $\beta_0(f)$ and then we deduce
$\lambda(\omega,T)$. $\beta_0(f)$ itself is in principle
temperature dependent, but we have checked that this temperature
dependence (which we may write to first order as 
$\beta(T)= \beta_0+\beta_1 T$) is negligible.
We have estimated $\beta_1 \sim 10^{-3} \beta_0\,\rm K^{-1}$.
This also implies that the choice of the reference temperature
where the correction factor is established is unimportant. 
Finally, the absolute value $\tilde{\lambda_0}$ is determined by
extrapolation to zero temperature of $\lambda(\omega,T)$.
\subsection{Estimate of the experimental uncertainty on $\tilde{\lambda_0}$}
\label{estimate}
The uncertainty on $\tilde{\lambda_0}$ comes firstly from the
experimental scatter (Fig.~\ref{fig1}).
From the set of different experiments, we have observed a scatter of
the normalized transmission of the order $\pm 40\%$ at 134 and 250$\,$GHz
and $\pm 25\%$ at 510$\,$GHz. This results roughly in a scatter of
$\pm 10\%$ for the penetration depth, hence $\pm \rm 200 \AA$ for a
typical value of 2000$\,$\AA. This uncertainty incorporates the error
of $\pm 100\,\rm \AA$ which we mentioned in the fitting to the
interference pattern.\\
A second type of uncertainty is a systematic error due to the value
of the resistivity $\rho_{110\,\rm K}$ at 110$\,$K, needed in the calculation
of the penetration depth. According to our own measurements (using the
Van Der Pauw technique) or other 4 points measurements requiring a
different geometrical factor, we estimate the error on $\rho_{110\,\rm K}$ 
to $\pm 10\%$ which typically leads to $\pm 5\%$ on the penetration
depth since it scales like $\rho_{110\,\rm K}^{1/2}$. For 2000$\,$\AA, it
represents $\pm 100\,\rm \AA$. This error is different from sample
to sample, but it is a systematic one. Moreover, it is smaller than
the  $\pm 200\,\rm \AA$ experimental uncertainty due to thermal drifts.
Finally, the absolute value $\tilde{\lambda_0}$ results from
an extrapolation at zero temperature. According to the temperature
variation which is observed, such an extrapolation cannot generate
an error larger than 40$\,$\AA. The absolute values for the penetration depth
will therefore be given within $\pm 200\,\rm \AA$.
\section{Samples}
\label{samples}
We present the results which we have obtained on three epitaxially grown,
laser ablated good quality YBCO thin films (referred as sample MA, MB,
and LR). Samples MA and MB are deposited on $\rm LaAlO_3$, and LR on MgO.
The substrates have a nominal $500\,\rm\mu m$ thickness and through
fitting, the thickness is found close to $510\,\rm \mu m$. The
characteristics of the samples are shown in Table~\ref{table1}.

The growth conditions of MA and MB samples have been described elsewhere
\cite{connell94}. The thickness of the films was not measured, but is
estimated to be 500$\,$\AA, based on measurements in similar samples which
had been patterned by photolithography. Not knowing precisely the
thickness is not a serious problem since it enters only as a correction
in the analysis of the transmission. Both samples display a resistivity
varying linearly with temperature, with close to zero extrapolation, and
a narrow transition (0.3$\,$K). These widths are indicative of good quality
samples, although the transition temperatures could be higher for samples
deposited onto  $\rm LaAlO_3$. Sample MB was then irradiated in order
to introduce point defects, following the conditions reported in a
systematic study of ion irradiated YBCO films \cite{connell94}.
After irradiation, the resistive transition of sample MB is lowered by
1.9$\,$K and slightly broadened (0.6$\,$K).\\
Sample LR was grown by laser ablation according to the method explained
in \cite{lepaventhivet95,castel95}. It has a narrow transition (0.8$\,$K), its
critical temperature is lower than the one obtained in samples grown
on $\rm LaAlO_3$ as usually observed for films grown on MgO. The narrow
rocking curve ($\rm \Delta\theta = 0.23^{\circ}$) is indicative of a
good $c$-axis orientation. Films grown on MgO often display simultaneous
in-plane epitaxial growth along the $[100]$ and $[1\overline{1}0]$
directions, which is very detrimental to surface resistance hence,
as was shown earlier, to the penetration depth \cite{lepaventhivet95}.
This specific sample was selected for its weak admixture of
$[1\overline{1}0]$ grains ($\rm 5\%$), giving rise to
$R_S(10\,\rm GHz, 77\, K) \simeq 2\, m\Omega$, deduced from the
experimental correlation between low frequency and microwave frequency
losses \cite{castel95}. This value is still much larger than what is now
commonly obtained in YBCO films, e.g.
$R_S(10\,\rm GHz, 77\, K) \leq 0.5\, m\Omega$.
\section{Penetration depth results}
\label{penetration}
We report now the penetration depth data derived as explained in the
previous section for the 3 samples examined in this paper.
Fig.~\ref{fig2} displays the results for sample MA (non irradiated,
$\rm LaAlO_3$). We have plotted the penetration depth for
$6\,{\rm K} \leq T \leq 60\,\rm K$ at five frequencies:
134, 250, 333, 387 and 510$\,$GHz.\\
The first parameter we extract is the value
$\tilde{\lambda_0} = 1990 \pm 200\,\rm \AA$ for this sample,
to be compared to the values found in single crystals: 1450$\,$\AA, from
muon spin rotation \cite{sonier94} or $1300\,\rm \AA$ (an average of the
penetration depth along the $a$ and $b$ directions measured by infrared
reflectivity \cite{basov95}), these values being believed to be close
to the intrinsic value.
Fig.~\ref{fig3a} and Fig.~\ref{fig3b} display the low temperature behavior,
$6\,{\rm K} \leq T \leq 30\,\rm K$, which is the one we focus on in
this study for the sake of the comparison to YBCO single crystals,
where the linear temperature dependence holds up to 30$\,$K. Fig.~\ref{fig3a}
is a plot versus $T$ and Fig.~\ref{fig3b} versus $T^2$. The data have
been systematically shifted by $20\,\rm \AA$ for clarity, starting from the
highest frequency. Whichever plot is considered, one notices immediately
that the temperature dependence of the penetration depth
$\lambda(\omega,T)$ becomes less and less pronounced as the
frequency increases. In this temperature range, we assign this behavior
to the frequency becoming comparable to the scattering rate
($\omega\tau \sim 1$) \cite{hirschfeld93b,hirschfeld94,bonn95,bontemps96}.
A similar qualitative change of the temperature dependence of
$\lambda(\omega,T)$ has been reported using two frequencies:
18.9$\,$GHz (cavity resonator technique) and 300$\,$GHz
(quasi-optical interferometer) \cite{dahne95}. Our study has the
advantage to use a single technique and five frequencies, therefore
allowing a more detailed analysis of the frequency effect on
$\lambda(\omega,T)$, as discussed below.\\
It is not possible to decide whether the temperature dependence is
linear or quadratic. The data are not accurate enough, but in any case,
as pointed out by Hirschfeld {\it et al.} \cite{hirschfeld93b,hirschfeld94},
once the temperature variation of the penetration depth is observed to
be frequency dependent, there is no analytic fit which would apply for
all frequencies. We have nevertheless quantified the effect of the
frequency on the penetration depth assuming either
$\lambda(\omega,T) = \tilde{\lambda_0} + \rm c_1(\omega) \it T$
(Fig.~\ref{fig3a}) or
$\lambda(\omega,T) = \tilde{\lambda_0} + \rm c_2(\omega) \it T^{\rm 2}$
(Fig.~\ref{fig3b}).
The value of the fitted parameters
$\rm c_1(\omega)$ and $\rm c_2(\omega)$
are reported in Table~\ref{table2}, within an error bar
of $\pm 0.7\,\rm \AA K^{-1}$ and $\pm 0.03\,\rm \AA K^{-2}$.\\

The results for sample LR (non irradiated, MgO) are shown in Fig.~\ref{fig4},
\ref{fig5a} and \ref{fig5b}. The extrapolated value of the penetration depth at
0$\,$K is $\tilde{\lambda_0} = 2180 \pm 200\,\rm \AA$.
There is a noteworthy difference when compared to sample MA, namely
the absence of a frequency dependence of $\lambda(\omega,T)$,
whether analyzed in terms of a $T$ or $T^2$ dependence:
$\rm c_1(\omega) \simeq 2.4 \pm 0.7\, \AA K^{-1}$,
$\rm c_2(\omega) \simeq 0.07 \pm 0.03\, \AA K^{-2}$ respectively.
For this sample, the experimental
scatter is larger than for sample MA or MB, because less experimental runs
have been performed at each frequency.
The absence of a clear temperature dependence implies that the
scattering rate is larger in this sample than in sample MA.\\ 

Fig.~\ref{fig6} displays the results in sample MB (irradiated,
$\rm LaAlO_3$) for three frequencies (134, 333 and 510$\,$GHz). We find
$\tilde{\lambda_0} = 2180 \pm 200\,\rm \AA$. 
In contrast with the two previous samples, the three curves in
sample MB have entirely collapsed up to 40$\,$K. The penetration depth
is independent of the frequency, which again can be understood
in terms of a large scattering rate ($\omega\tau \ll 1$). Moreover,
the deviations for $T > 40\,\rm K$ are almost negligible.
Secondly, in contrast with the two other samples, the inset of
Fig.~\ref{fig6} shows a clear $T^2$ dependence, holding up to 40$\,$K.
$\rm c_2(\omega) = 0.07 \pm 0.03\, \AA K^{-2}$.\\
Overall, samples MB and LR appear to bear some similarity,
somewhat surprisingly given the fact that they are actually different,
whereas sample MA exhibits a  shorter penetration depth and a significant
effect of the frequency.
\section{Analysis}
\label{analysis}
We now turn to a more quantitative analysis in the framework of the
two fluids model, which makes use of a Drude conductivity for the
normal fluid, involving a frequency independent scattering rate
$1/\tau$. This model has been widely used in order to analyze the
conductivity, whether in the microwave range
\cite{shibauchi96,bonn94,bonn93,bonn93b,shibauchi92}
or the THz regime \cite{dahne95,frenkel96}.
One reason for using it is that it yields the right trends.
The maximum in the real part $\sigma_1(\omega,T)$
of the conductivity below $T_c$, observed in single crystals
\cite{bonn94,bonn93,bonn93b,shibauchi92}
and in thin films \cite{hensen97,farber,dahne95,nuss91,frenkel96}
can be interpreted within the two fluids model, as a result of the
rapid decrease of the scattering rate $1/\tau$ below $T_c$.
It is found experimentally that this maximum of $\sigma_1(\omega,T)$
shifts to higher temperatures, and its magnitude decreases as the
frequency increases. This is understood within the same framework,
where the maximum is shown to occur when $\omega \tau \sim 1$.
Meanwhile, the two fluids model is also widely recognized as being
inappropriate. Such a simple phenomenology does not describe properly
$\sigma_1(\omega,T)$ in BCS superconductors.
In high-$T_c$ superconductors, it does not yield a quantitatively
correct description of the conductivity and of the scattering rate as
a function of frequency and/or temperature. From the experimental
point of view, there are indeed indications that the scattering rate
is frequency dependent \cite{shibauchi96,nuss91,frenkel96}.
On the theoretical side, the conductivity
$\sigma_1(\omega,T)$ has been evaluated by Hirschfeld {\it et al.}
for unconventional $d$-wave superconductors \cite{hirschfeld93b,hirschfeld94}.
In the so-called pure regime and for a very small concentration of
impurities, a Drude-like conductivity is recovered however the
scattering rate is frequency dependent ($1/\tau(\omega)$)
\cite{hirschfeld93b,hirschfeld94,hensen97}. Under specific assumptions,
$1/\tau(\omega)$ may not depend upon the frequency, but the whole
calculation does not provide an actual support for the use of the
standard two fluids model where $1/\tau$ depends only on temperature.
Interestingly, the change of $\omega\tau$ from
$\omega\tau \ll 1$ to  $\omega\tau \gg 1$ yields a definite change
in the temperature dependence of $\lambda(\omega,T)$ at low temperature. 
For the sake of the comparison with earlier data in the literature,
and also in order, as mentioned above, to establish the order of
magnitude for  $1 /\tau$  from the penetration depth data,
we nevertheless perform the analysis of our data using this phenomenology. 
The change in the penetration depth
$\lambda(\omega,T)$ with the frequency (from 100$\,$GHz to 500$\,$GHz)
was described in an earlier paper within the assumption of a linear
behavior of $\lambda(\omega,T)$ \cite{bontemps96}.
The calculation of $\lambda(\omega,T)$ used the two fluids model
for three frequencies 130, 330, 510$\,$GHz and considered various
elastic scattering rates
$1/\tau_{el}$, each one combining according to Mathiessen law with
the inelastic scattering rate $1/\tau_{inel}$, the latter being
assumed to be adequately described by the data taken from single crystals
\cite{bonn94,bonn93,bonn93b}. We recall briefly how the calculation
was performed when applied to $\lambda(\omega,T)$ and not to
$\sigma_1(\omega,T)$ as usual: The superfluid and normal fluid fractions are
$x_s(T)$ and $x_n(T)$ respectively, so that $x_s(T)+x_n(T) = 1$;
$x_s(T) = \tilde{\lambda_0^2} /\lambda^2(\omega=0,T)$.
One assumes a Drude form for the frequency dependence of the normal
fluid conductivity. Thus, the real and imaginary part of the total
conductivity write:

\begin{eqnarray}
{\sigma_1(\omega,T)} = {1 \over \mu_0 \omega \lambda_0^2}\
x_n(T)\ {\omega \tau \over 1 + \omega^2 \tau^2}
\label{eq5}
\end{eqnarray}

\begin{eqnarray}
{\sigma_2(\omega,T)} = {1 \over \mu_0 \omega \lambda_0^2}
\Biggl [x_n(T) {\omega^2 \tau^2 \over 1 + \omega^2 \tau^2} + x_s(T)\Biggl]
\label{eq6}
\end{eqnarray}

Note that (\ref{eq6}) gives an explicit expression of
$\lambda(\omega,T)$, which was introduced more formally in (\ref{eq3}).
The two extreme cases are now easily understood: 
for $\omega\tau \ll 1$ (quasi-static regime), the first term of
$\sigma_2(\omega,T)$ is negligible at low temperatures and
$\lambda(\omega,T)$ is just the frequency independent London penetration depth
$\lambda_L(T)$. For $\omega\tau \gg 1$ (collisionless regime),
$\lambda(\omega,T) = \lambda_0$ = constant, so that the
penetration depth is both frequency and temperature independent,
as indeed observed in the infrared range above $20\,\rm cm^{-1}$
(600$\,$GHz) \cite{basov95}.
The underlying physics is that, as $\omega\tau$ increases,
the inductive contribution of the normal fluid increases and yields
an additional screening of the electromagnetic field, thus a smaller
increase of the penetration depth with temperature, the maximum possible
screening corresponding to $\lambda_0$.
Note that this caculation cannnot explain an increased penetration depth
$\tilde{\lambda_0}$.

We may now come back to Table~\ref{table2}, which displays for sample MA the
values of the parameters $\rm c_n\it (\omega)$ (n=1 and n=2) where
$\lambda(\omega,T) = \tilde{\lambda_0} + \rm c_n(\omega)\it T^{\rm n}$.
Assuming a constant scattering rate at low temperatures ($T < 30\,\rm K$,
not inconsistent with the published results) and using (\ref{eq6}), the two
fluids model yields the frequency dependence of $\rm c_n(\omega)$
{\it whatever} n:
 
\begin{eqnarray}
{\rm c_n(\omega)} = {\rm c_n(0) \over {1+\omega^2 \tau^2}}
\label{eq7}
\end{eqnarray}

Fig.~\ref{fig7a} and \ref{fig7b} show a fit to (\ref{eq7}) of the experimental
data from sample MA for $\rm c_n$ (n=1 and n=2, see Table~\ref{table2}) versus
the frequency. Two fitting parameters are required:
$\rm c_n(0)$ and $1/\tau$. For the sake of internal consistency,
the scattering rate deduced from the fit for n=1 and n=2 must be the
same. We find indeed the best agreement in both cases for
$1/\tau = 1.7 \times 10^{12}\,\rm s^{-1}$.
As a comparison, Bonn {\it et al.} \cite{bonn93}, estimating $1/\tau$
for YBCO single crystals within the two fluids model, obtain a value
of $1/\tau =  5 \times 10^{11}\,\rm s^{-1}$, 3 times smaller than the value
we find. Our result is closer to the value $1.4 \times 10^{12}\,\rm s^{-1}$
found for a $0.31\, \%$ Zn-doped single crystal
\cite{bonn93b}, where the authors show that adding Zn impurities
increases the low temperature limit of the scattering rate.
Therefore the scattering rate in sample MA appears to be closer to the
one of a single crystal with impurities, which confirms the common
idea that the films contain more defects than the single crystals.
The nature of these defects remains to be identified, as they seem to
affect very differently the absolute value, the temperature variation
and the frequency dependence of $\lambda(\omega,T)$.
We will discuss the two latter topics further.\\
The second parameter we obtain from the fit is the slope at zero
frequency $\rm c_n(0)$. For n=1,
$\rm c_1(0) = 3.5 \pm 0.7\, \AA K^{-1}$. It is compatible with
the canonical value $\simeq 4\,\rm \AA K^{-1}$ found in single crystals
\cite{hardy93}. For n=2, $\rm c_2(0) = 0.10 \pm 0.03\, \AA K^{-2}$;
we will use this in the next section.\\
The scattering rate for sample LR
is not measurable in our experimental set-up since the temperature
variation of the penetration depth is not frequency dependent
within our experimental accuracy. However from a putative fit to (\ref{eq7})
compatible with the experimental uncertainty, we can estimate a
lower bound of $3 \times 10^{12}\,\rm s^{-1}$.\\
MB is a somewhat special sample since point defects have been introduced
on purpose by irradiation. This sample was similar to sample MA before
irradiation (same growth conditions, same resistivity, same $T_c$).
The irradiation lowered $T_c$ down to 87.8$\,$K, and raised the
110$\,$K resistivity to $105\,\rm \mu\Omega cm$. The defects introduced
by irradiation are thought to act as extra scattering centers.
This is roughly consistent with the resistivity data.
The normal state resistivity after irradiation has increased, and it is
still linear with temperature; the slope is however slightly different
from the one of the pristine sample, contrary to what is observed in
similarly irradiated films \cite{connell94} and electron irradiated
crystals \cite{giapintsakis94}.
In order to check the consistency of our data with the condition
$\omega\tau \ll 1$ expected to be fulfilled in this sample,
we do the following estimate. In single crystals, above $T_c$,
a typical $80\,\rm \mu\Omega cm$ resistivity corresponds to an
inelastic scattering
rate $1/\tau_{inel} = 3 \times 10^{13}\,\rm s^{-1}$ \cite{bonn94}.
The increase of resistivity at 110$\,$K
introduced by irradiation is then equivalent to adding
$1/\tau_{el} \simeq 9 \times 10^{12}\,\rm s^{-1}$.
A calculation in the two fluids framework shows
indeed that taking $1/\tau_{el} \simeq 10^{13}\,\rm s^{-1}$
yields no frequency dependence in
$\lambda(\omega,T)$ within our experimental resolution.
This also explains qualitatively why the penetration depth in
sample MB hardly shows any frequency dependence even for
$T > 40\,\rm K$ (Fig.~\ref{fig6}).\\ 

We have shown in this section that the two fluids model allows
a reasonable understanding of the experimental data if not yielding
the exact figures. The underlying physics in (\ref{eq7}), whose simplicity
derives of course directly from the two fluids model, is that the
measuring frequency is comparable to the scattering rate. Whichever
specific model we may use, we expect that when these two figures
are of the same order of magnitude, the experimentally measured
penetration depth must be modified, which is indeed what is derived
in \cite{hirschfeld93b,hirschfeld94}.
\section{Discussion of the d-wave model and the resonant scattering}
\label{discussion}
None of our samples exhibit an optimal $T_{c0} = 92\,\rm K$.
Our penetration depth data are not accurate enough so as to make a
clear distinction between a $T$ or $T^2$ dependence at
$T \leq 30\,\rm K$.
In this temperature range, the temperature dependence for
$\lambda(\omega=0,T)$ in sample MA is consistent with
the standard linear behavior ($\sim 4\,\rm \AA K^{-1}$). The two other
samples exhibit a slower increase of $\lambda(\omega=0,T)$
with temperature: this appears when comparing for instance the increase of
$\lambda(\omega=0,T)$ which is characterized in the first
column of Table~\ref{table3} by the parameter $\rm c_2(0)$.\\ 
All the samples display a penetration depth $\tilde{\lambda_0}$
significantly increased with respect to the generally accepted value
$\lambda_0 = 1400 \pm 100\,\rm \AA$ \cite{sonier94,basov95}.
In this section, we analyze the data in the framework of the $d$-wave model
with impurity scattering, somewhat similarly to what was done in single
crystals and thin films \cite{hensen97,ma93,bonn94,ulm95}.
The noticeable difference with earlier work is that we may now use a
reliable and more accurate measurement of $\tilde{\lambda_0}$
to estimate the change $\delta\lambda_0 = \tilde{\lambda_0} - \lambda_0$.
We proceed as follows: $\delta\lambda_0$ is related to the parameter
$\rm c_2(0)$ which controls the variation of the penetration depth
in the gapless regime \cite{hirschfeld93,hirschfeld93b,hirschfeld94}. From
$\rm c_2(0)$, we deduce the cross-over temperature where the linear
regime is recovered, $T^*$, using $\lambda_0 = 1400\,\rm \AA$\ and
$\Delta_0 = \lambda_0 ln2 / \rm c_1(0) = 243\, K$
(from the standard value $\rm c_1(0) = 4\,\AA K^{-1}$
in crystals). This yields the expected change $\delta \lambda_0$ in the
framework of the model, assuming the unitary limit, which we compare to
our experimental estimate. The values of $\rm c_2(0)$, $T^*$,
$\delta\lambda_0$ (theoretical) and $\delta\lambda_0$ (experimental)
are displayed in the first four columns of Table~\ref{table3}. Earlier work
skipped the estimate of $T^*$ by plotting $\delta\lambda_0$ versus $\rm c_2(0)$,
or a similar parameter association. It is instructive though to quote $T^*$,
especially in our case where the comparison shows clearly that the
experimental values of  are much larger than the ones deduced from
$T^*$. Indeed, $T^*$ (which is directly related to the
scattering rate, hence the impurity concentration) is anomalously large:
one would expect $T^{*} < 30\,\rm K$, since the model predicts
that the linear regime is recovered above $T^*$, and in YBCO,
this linear regime only holds up to 30~K. This may point one possible
reason why the model does not work, before ruling it out, namely the
amount of impurities or defects in the films is too large and the model
does no longer apply. The last two columns of Table~\ref{table3} confirm the
failure of the model to account for the decrease of the critical
temperature related to the decrease of the superfluid density, hence
the increase of $\lambda_0$. We deduce from the model \cite{kim94}
the reduction of $T_c$ ($T_c/T_{c0}$--theoretical)
associated to $\delta\lambda_0$ and we compare it to the actual ratio,
($T_c/T_{c0}$--experimental). Again, this comparison shows that
even in the unitary limit, the increase of $\lambda_0$ should be
associated with a larger decrease of the critical temperature.\\
Although it is commonly quoted that the films may bear some similarity
with the impurity doped single crystals, our data compared with the
theoretical expectations in the $d$-wave model show that this similarity
is far from being obvious, once the absolute value of $\tilde{\lambda_0}$
is determined.\\

Our discussion is of course only valid if we assign the increase of the
penetration depth with respect to the intrinsic value entirely to
resonant scattering due to impurities. It is however quite possible that this
increase may be
related to other very different defects. To illustrate this point,
we discuss briefly the example of extended defects such as grain
boundaries, likely to be present in thin films.
Grain boundaries which form resistively shunted Josephson junctions
with Josephson penetration depth  $\lambda_J$ lead to an effective
penetration depth $\lambda_{eff}$ which reads \cite{hylton88}:
$\lambda_{eff}^2 = \tilde{\lambda_0^2} + \lambda_J^2$.  
The effect of the grain boundaries is then to enhance the intrinsic
penetration depth, which would provide an explanation for the large
values that we find. A first question which arises immediately is
whether such grain boundaries may be effective for scattering.
Recall indeed that we find evidence for a scattering rate 
$1/\tau = 1.7 \times 10^{12}\,\rm s^{-1}$ in sample MA,
which suggests a good quality, in contrast with the large penetration
depth. With a Fermi velocity $v_F = 2 \times 10^7\,\rm cm s^{-1}$ 
\cite{allen88}, the QP mean free path $\ell$ at low temperature would be
$\ell = v_F \tau = 1300\,\rm \AA$, which is roughly the size
of the grains \cite{perrin}. Hence grain boundaries may be weakly
scattering, which does not contradict their 
being responsible for a large penetration depth. This assumption does
not however look acceptable since all the films under investigation
exhibit standard critical current densities, which imply firstly
presumably a short $\lambda_J$, secondly an extra temperature variation
due to $\lambda_J$. Indeed, the Josephson critical current in an
unconventional superconductor with nodes in the gap is expected to
display a definite temperature dependence at low temperature
(unlike the standard BCS Ambegaokar-Baratoff expression) \cite{burkhardt}.
This yields a more complicated temperature dependence for
$\lambda_{eff}$, which must be difficult to identify.\\
Another explanation could be the occurrence of pinholes in the films.
However, putting numbers rules out such a possibility: a typical amount
of pinholes would give rise at most to $10\, \%$ leakage \cite{moffat}.
Such a leakage results in an increase of $\sim 50\,\rm \AA$ for the
penetration depth, and we are looking for an increase of
$\sim 500\,\rm \AA$ (see $\delta\lambda_0$--experimental in
Table~\ref{table3}).\\

We finally turn toward the possible validity of the two fluids model
in the framework of the $d$-wave model \cite{hirschfeld93b,hirschfeld94}. For
$T > T^*$, the two fluids expression for the conductivity
is Drude-like, except that the scattering rate is now frequency dependent.
Is is shown however that there is a frequency range, namely
$\omega < T^*$, where $1/\tau$ is constant and its value is then
$2 T^*$. Let us examine whether this may apply in our case.
We find in sample MA $1/\tau = 1.7 \times 10^{12}\,\rm s^{-1} = 13\, K$,
hence in this case we would expect
$T^{*} = 1 / 2\tau \simeq 6\,\rm K$. The smallest frequency
we use is 130$\,$GHz also of the order of 6$\,$K. Hence, our experimental
conditions do not meet the requirement of the model, which then
predicts a frequency dependence of the scattering rate.
\section{Conclusion}
\label{conclusion}
We have established in a pristine sample YBCO/$\rm LaAlO_3$, a definite
effect of the
frequency on the measured penetration depth, e.g. the low temperature
variation gets smaller as the frequency increases
(a similar effect has been observed in two other samples from a
different origin). The discussion of the actual temperature dependence
of the penetration depth $\lambda(\omega,T)$ has been purposely
skipped here and is postponed to  a further report.
The frequency effect can be consistently described by the two
fluids model, which yields the low temperature scattering rate.
However, this two fluis analysis has so far no reliable theoretical background
for unconventional superconductors. We have stressed that the
evolution of this temperature dependence with frequency
results more generally from a competition between the scattering rate
and the frequency. In this respect, our experiment puts  a reasonable
order of magnitude for the quasi-particle scattering time in thin films
at low temperature.\\
Our comparison to the $d$-wave model shows that this model fails to
account for the various parameters which are measured and which
are expected to be related through the theory. We note that this failure
does not necessarily rule out the validity of the model.
Indeed, our final discussion raises several caveat: firstly, although
our samples look as good if not better than the ones referred to in the
literature, they may still include too many defects, which then puts
them out of the range of validity of the model. Secondly, although we
insist on the essential knowledge of $\tilde{\lambda_0}$, an
increased penetration depth
may result from extrinsic sources yet to be identified, in which case
its value cannot be taken for the comparison with the theory. Therefore
the major point which we eventually wish to strongly emphasize is that
any comparison with a theoretical model, if dealing with thin films,
should be considered with extreme care. Even in experimentally
satisfactory conditions, where a fairly complete set of parameters
is available, such a comparison is still ambiguous.

\begin{acknowledgement}
We are very grateful to G.~Deutscher and P.~Monod for many illuminating
discussions and helpful suggestions.
\end{acknowledgement}

\bibliographystyle{prsty}
\bibliography{biblio}

\onecolumn

\begin{figure}
\caption{Normalized transmission $T_{10\,\rm K}/T_{110\,\rm K}$ versus the
square of the frequency in sample MA. For a same nominal frequency, each point
corresponds to an experimental run, after which the sample is dismounted before
running the measurement again. The interference pattern (solid line) is
calculated \protect\cite{glover57} using $l=515\,\mu\rm m$, $n=5$ 
and $\lambda(10\,\rm K) = 2000\,\AA$ in order to fit the experimental data.
The dashed line represents (\ref{eq4}), e.g. the simple expression for the
normalized transmission, neglecting the interference pattern, for the same
parameters $n$ and $\lambda(10\,\rm K)$.}
\label{fig1}
\end{figure}

\begin{figure}
\caption{Penetration depth for sample MA versus temperature for
$T \le 60\,\rm K$ at various frequencies: 134, 250, 333, 387, 510$\,$GHz. Above
60$\,$K, the derivation of $\lambda(\omega,T)$ is no longer valid.
$\tilde{\lambda_0}$(extrapolated)=1990$\,$\AA. The frequency dependence of the
penetration depth below $T \le 30\,\rm K$ is clearly amplified above this
temperature.}
\label{fig2}
\end{figure}

\begin{figure}
\caption{Penetration depth for sample MA versus temperature $T$ for
$T \le 30\,\rm K$ (where a linear regime is commonly observed in single
crystals)
for five frequencies: 134, 250, 333, 387, 510$\,$GHz. Best fits to a linear
variation $\lambda(\omega,T) = \tilde{\lambda_0} + \rm c_1(\omega) \it T$ are
displayed for each curve as well as the value $\rm c_1(\omega)$ of the slope.
The curves are systematically shifted up by 20$\,$\AA\ for clarity, starting
from the highest frequency. This figure shows clearly the change of the
temperature dependence with increasing frequency.}
\label{fig3a}
\end{figure}

\begin{figure}
\caption{Same data as in Fig.~\ref{fig3a} plotted versus $T^2$. Best fits to a
$T^2$ variation
$\lambda(\omega,T) = \tilde{\lambda_0} + \rm c_2(\omega) \it T^{\rm 2}$
are displayed for each curve, as well as the value of
$\rm c_2(\omega)$. Within the accuracy of the measurement, we cannot really
distinguish between a $T$ or $T^2$ dependence.}
\label{fig3b}
\end{figure}

\begin{figure}
\caption{Penetration depth for sample LR versus temperature for
$T \le 60\,\rm K$ at five frequencies: 134, 250, 333, 387, 510$\,$GHz. Above
60$\,$K, the derivation of $\lambda(\omega,T)$ is no longer valid.
$\tilde{\lambda_0}$(extrapolated)=2180$\,$\AA. There is no frequency dependence
for $T \le 30\,\rm K$, but the curves are clearly separated from each other
above $\sim 40\,\rm K$.}
\label{fig4}
\end{figure}

\begin{figure}
\caption{Penetration depth from sample LR versus temperature $T$ for
$T \le 30\,\rm K$ (where a linear regime is commonly observed in single
crystals) for five frequencies: 134, 250, 333, 387, 510$\,$GHz. Best fits to a
linear variation
$\lambda(\omega,T) = \tilde{\lambda_0} + \rm c_1(\omega) \it T$ are
displayed for each curve as well as the value $\rm c_1(\omega)$ of the slope.
The curves are systematically shifted up by 50$\,$\AA\ for clarity. The average
slope is $2.4\,\rm \AA K^{-1}$.}
\label{fig5a}
\end{figure}

\begin{figure}
\caption{Same data as in Fig.~\ref{fig5a} plotted versus $T^2$. Best fits to a
$T^2$ variation variation
$\lambda(\omega,T) = \tilde{\lambda_0} + \rm c_2(\omega) \it T^{\rm 2}$
are displayed for each curve, as well as the value of
$\rm c_2(\omega)$. The average slope is $0.064\,\rm \AA K^{-2}$. As
in sample MA, within the accuracy of the measurement, we cannot really
distinguish between a $T$ or $T^2$ dependence.}
\label{fig5b}
\end{figure}

\begin{figure}
\caption{Penetration depth for sample MB versus temperature for
$T \le 60\,\rm K$ at three frequencies: 134, 333, 510$\,$GHz. Above 60$\,$K,
the derivation of $\lambda(T)$ is no longer valid.
$\tilde{\lambda_0}$(extrapolated)=2180$\,$\AA. There is no frequency dependence
for $T \le 30\,\rm K$. The inset shows the data plotted versus $T^2$. In
contrast with MA and LR, the curves collapse in a large temperature range
typically up to 50$\,$K.}
\label{fig6}
\end{figure}

\begin{figure}
\caption{Experimental data $\rm c_1(\omega)$ from
$\lambda(\omega,T) = \tilde{\lambda_0} + \rm c_1(\omega) \it T$ at five
frequencies from the data shown in Fig.~\ref{fig3a} along with the fitting
parameters $1/\tau$ and $\rm c_1(0)$.
$1/\tau = 1.7 \times 10^{12}\,\rm s^{-1}$.
$\rm c_1(0) = 3.5 \pm 0.7\, \AA K^{-1}$, consistent with the
canonical slope found in single crystals.}
\label{fig7a}
\end{figure}

\begin{figure}
\caption{Experimental data $\rm c_2(\omega)$ from
$\lambda(\omega,T) = \tilde{\lambda_0} + \rm c_2(\omega) \it T^{\rm 2}$ at five
frequencies from the data shown in Fig.~\ref{fig3b} along with the fitting
parameters $1/\tau$ and $\rm c_2(0)$. The scattering rate $1/\tau$ is the
same as in Fig.~\ref{fig7a}.}
\label{fig7b}
\end{figure}

\onecolumn
\begin{table}
\caption{Characteristics of the three samples MA, MB and LR: prep (growth
 technique), critical temperature $T_c$ and resistive width
$\rm\Delta\it T_c$, thickness $d$ of the film, thickness $l$ of the substrate,
width $\rm\Delta\theta$ of the rocking curve (when available), surface
resistance $R_S$ at 10$\,$GHz and 77$\,$K (when available), irradiation dose
by He ions.}
\label{table1}
\begin{tabular}{cccccccccc}
\hline\noalign{\smallskip}
Sample & prep. & $T_c(R=0)$ & $\rm\Delta\it T_c$ & $d$ & $\rho_{110\,\rm K}$ &
$l$ & $\rm\Delta\theta$ & $R_S$ & He\\
&& (K) & (K) & (\AA) & ($\rm\mu\Omega cm$) & ($\mu$m) & & (m$\rm\Omega$) &
($10^{14}\,\rm He/cm^2$)\\
\noalign{\smallskip}\hline\hline\noalign{\smallskip}
MA & laser & 90.2 & 0.3 & 500 & 90 & 515 & & & none\\
\noalign{\smallskip}\hline\noalign{\smallskip}
MB & laser & 89.7 & 0.3 & 500 & 80 & 510 & & & none\\
& after irr. & 87.8 & 0.6 & & 105 & & & & 2\\
\noalign{\smallskip}\hline\noalign{\smallskip}
LR & laser & 88.5 & 0.8 & 1300 & 100 & 515 & $0.23\,^{\circ}$ & 2 & none\\
\noalign{\smallskip}\hline
\end{tabular}
\end{table}

\begin{table}
\caption{$\rm c_1(\omega)$ and $\rm c_2(\omega)$ for samples MA and LR at the
various frequencies. The parameters $\rm c_1(\omega)$ and $\rm c_2(\omega)$ are
defined through the fitting of the penetration depth (shown in
Fig.~\ref{fig3a}, \ref{fig3b}, \ref{fig5a} and \ref{fig5b}) to
$\lambda(\omega,T) = \tilde{\lambda_0} + \rm c_1(\omega) \it T$  and
$\lambda(\omega,T) = \tilde{\lambda_0} + \rm c_2(\omega) \it T^{\rm 2}$
respectively. The uncertainty is $\pm 0.7\,\rm \AA K^{-1}$ on $\rm c_1(\omega)$
and $\pm 0.03\,\rm \AA K^{-2}$ on $\rm c_2(\omega)$.}
\label{table2}
\begin{tabular}{c|cc|cc}
\hline\noalign{\smallskip}
&
\multicolumn{2}{c}{
   \begin{tabular}{c}
   \bf Sample MA
   \end{tabular}
} &
\multicolumn{2}{c}{
   \begin{tabular}{c}
   \bf Sample LR
   \end{tabular}
}\\
Frequency & $\rm c_1 exp(\omega)$ & $\rm c_2 exp(\omega)$ &
$\rm c_1 exp(\omega)$ & $\rm c_2 exp(\omega)$\\[3pt]
(GHz) & (\AA K$^{-1}$) & (\AA K$^{-2}$) & (\AA K$^{-1}$) & (\AA K$^{-2}$)\\
\noalign{\smallskip}\hline\hline\noalign{\smallskip}
134 & 2.7 & 0.073 & 3.1 & 0.086\\
250 & 1.8 & 0.050 & 3.3 & 0.096\\
333 & 1.4 & 0.040 & 1.6 & 0.043\\
387 & 1.2 & 0.033 & 2.2 & 0.054\\
510 & 0.8 & 0.022 & 2.0 & 0.058\\
\noalign{\smallskip}\hline
\end{tabular}
\end{table}

\begin{table}
\caption{Comparison of the experimental: $\delta\lambda_0$--exp, and
theoretical $\delta\lambda_0$--theo, changes of the penenration depth deduced
from the $\rm c_2$ parameter \protect\cite{hirschfeld93}. $T^*$ is the
cross-over temperature, also deduced
from $\rm c_2$. The table shows also the experimental and theoretical relative
change of the critical temperature ($T_c/T_{c0}$) in the framework of the
$d$-wave model with impurity scattering \protect\cite{kim94} (see text).}
\label{table3}
\begin{tabular}{ccccccc}
\hline\noalign{\smallskip}
& $\rm c_2$ & $T^*$ & $\delta\lambda_0$ & $\delta\lambda_0$ & $T_c/T_{c0}$
& $T_c/T_{c0}$\\
&&& theo & exp & theo & exp\\
& (\AA K$^{-2}$) & (K) & (\AA) & (\AA)\\
\noalign{\smallskip}\hline\hline\noalign{\smallskip}
MA & 0.095 & 42 & 195 & 590 & 0.75 & 0.98\\
LR & 0.07 & 60 & 254 & 780 & 0.6 & 0.95\\
MB & 0.07 & 59 & 252 & 780 & 0.6 & 0.96\\
\noalign{\smallskip}\hline
\end{tabular}
\end{table}
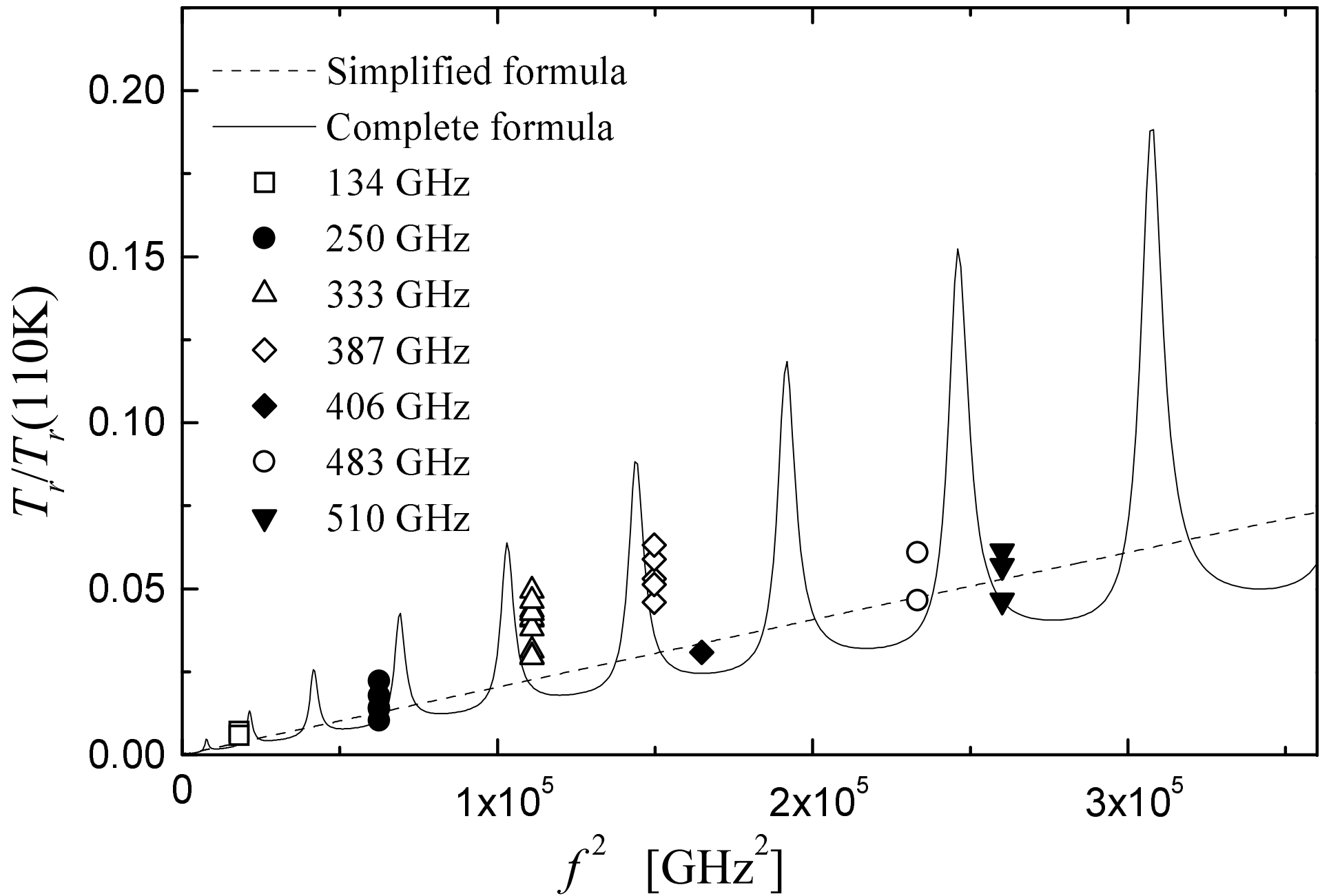
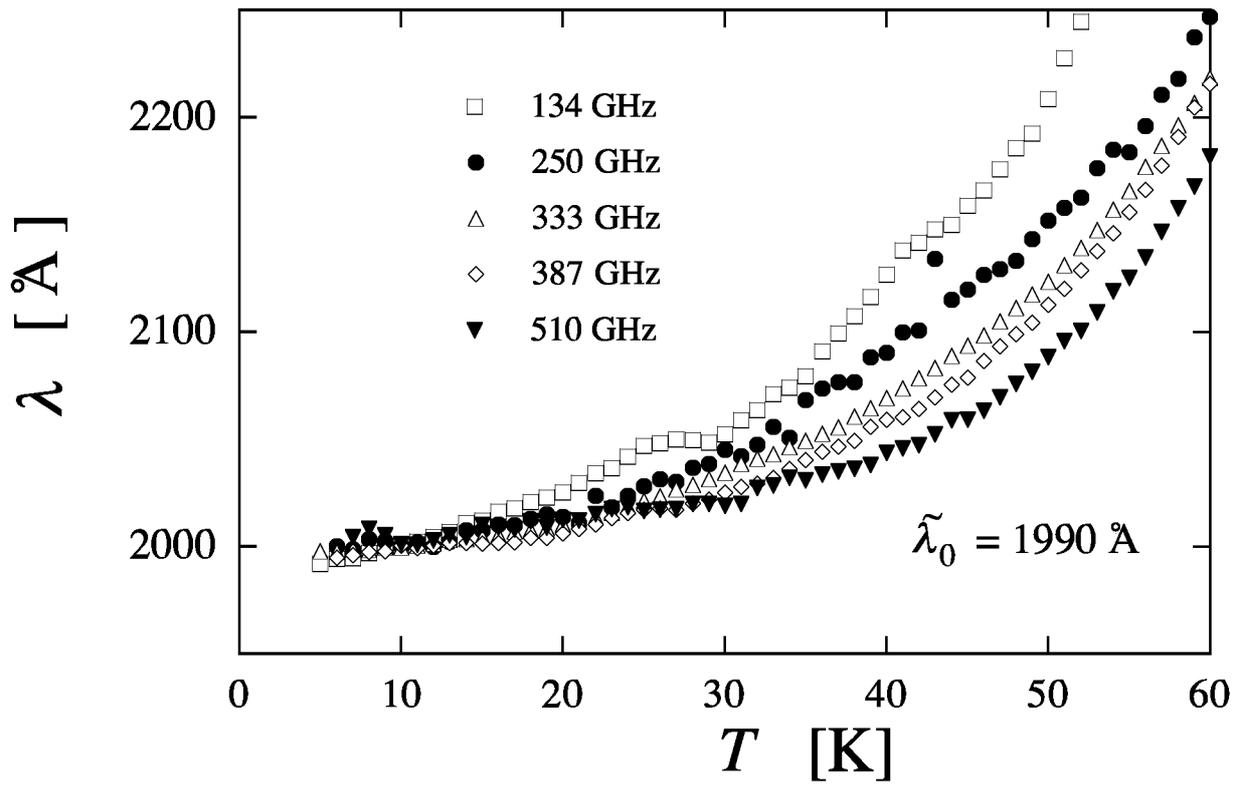
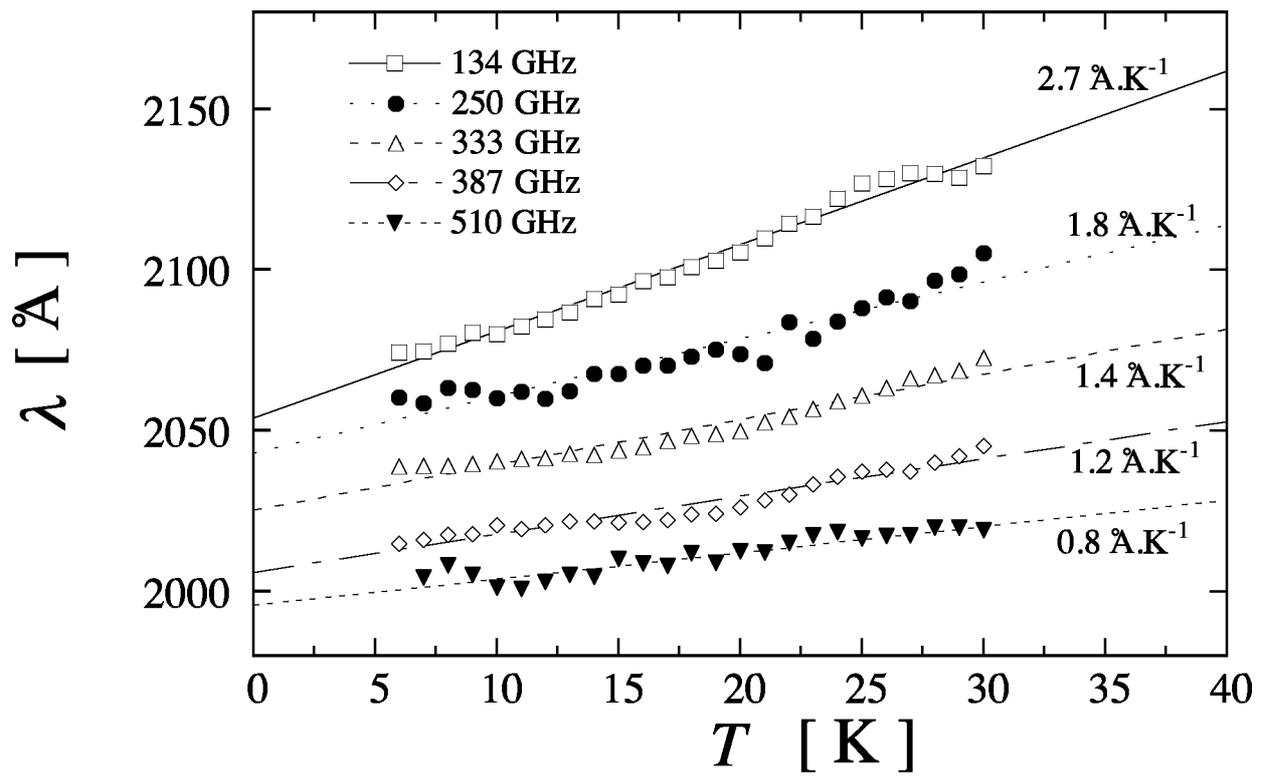
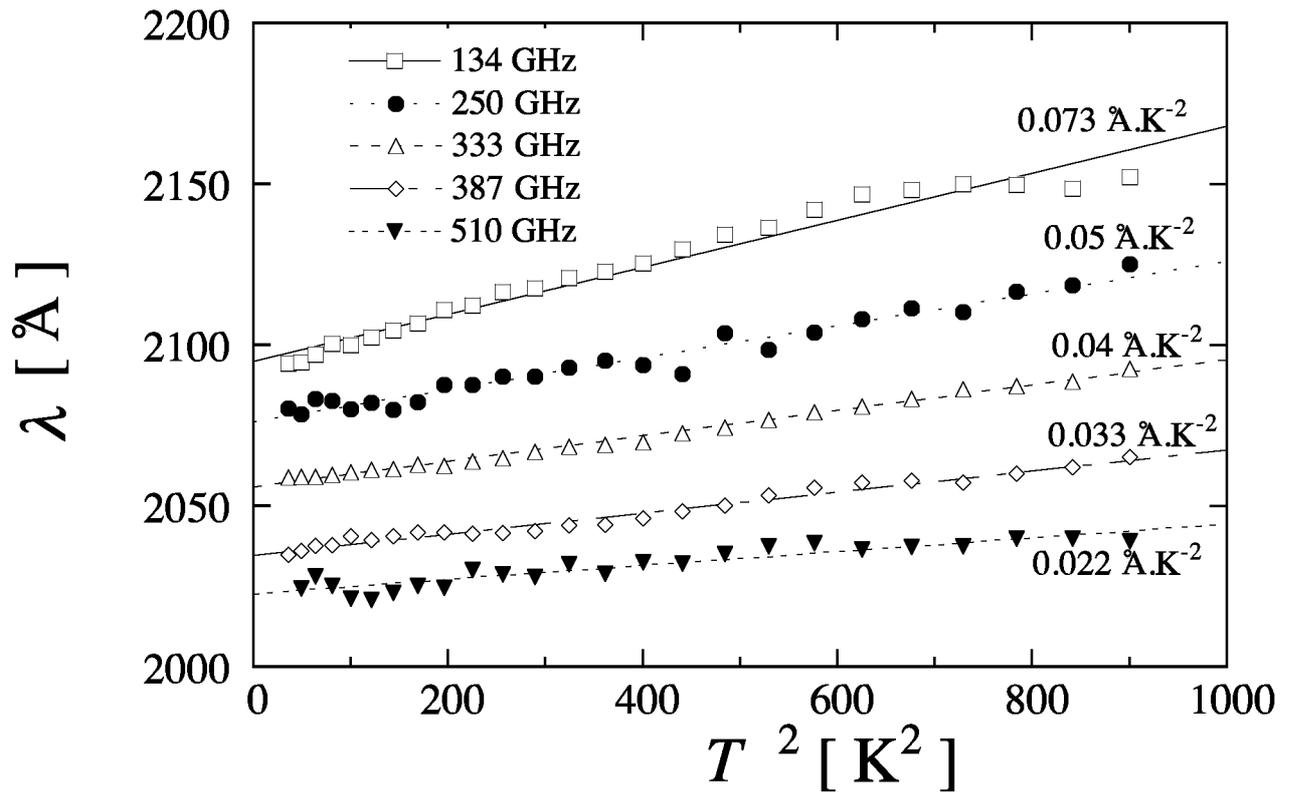
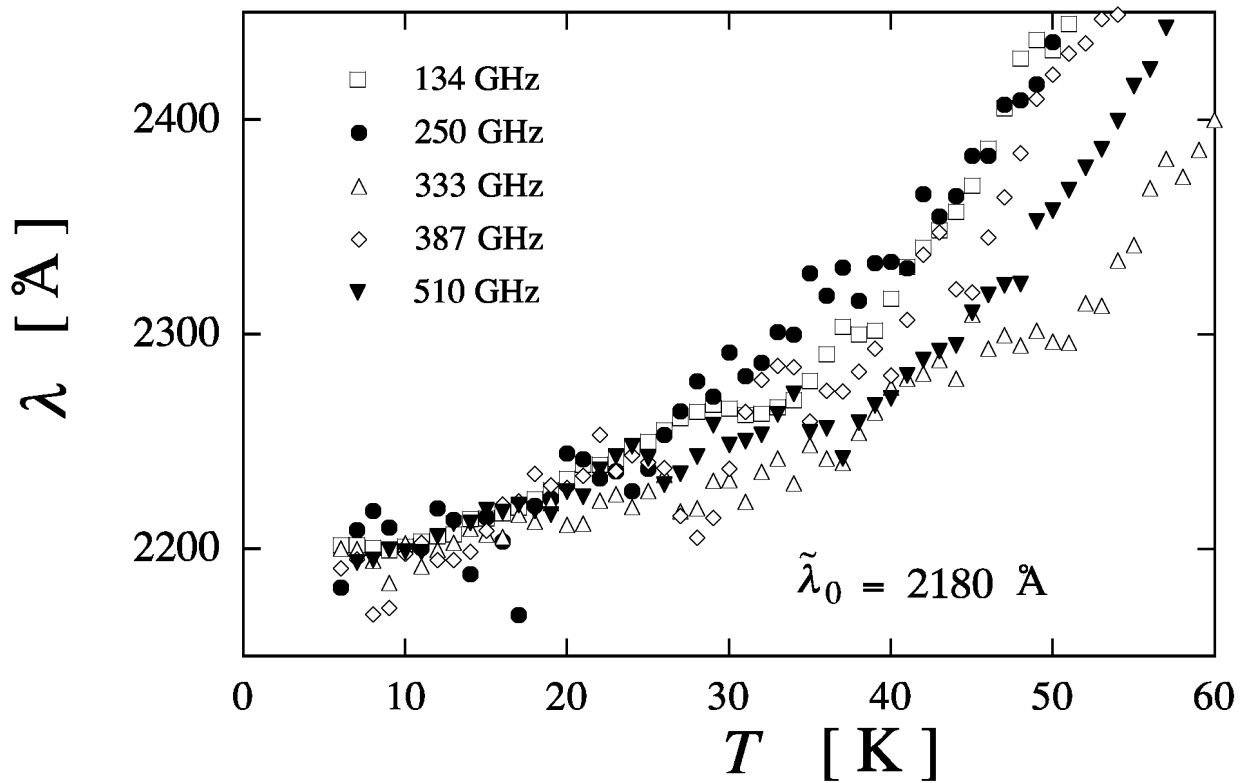
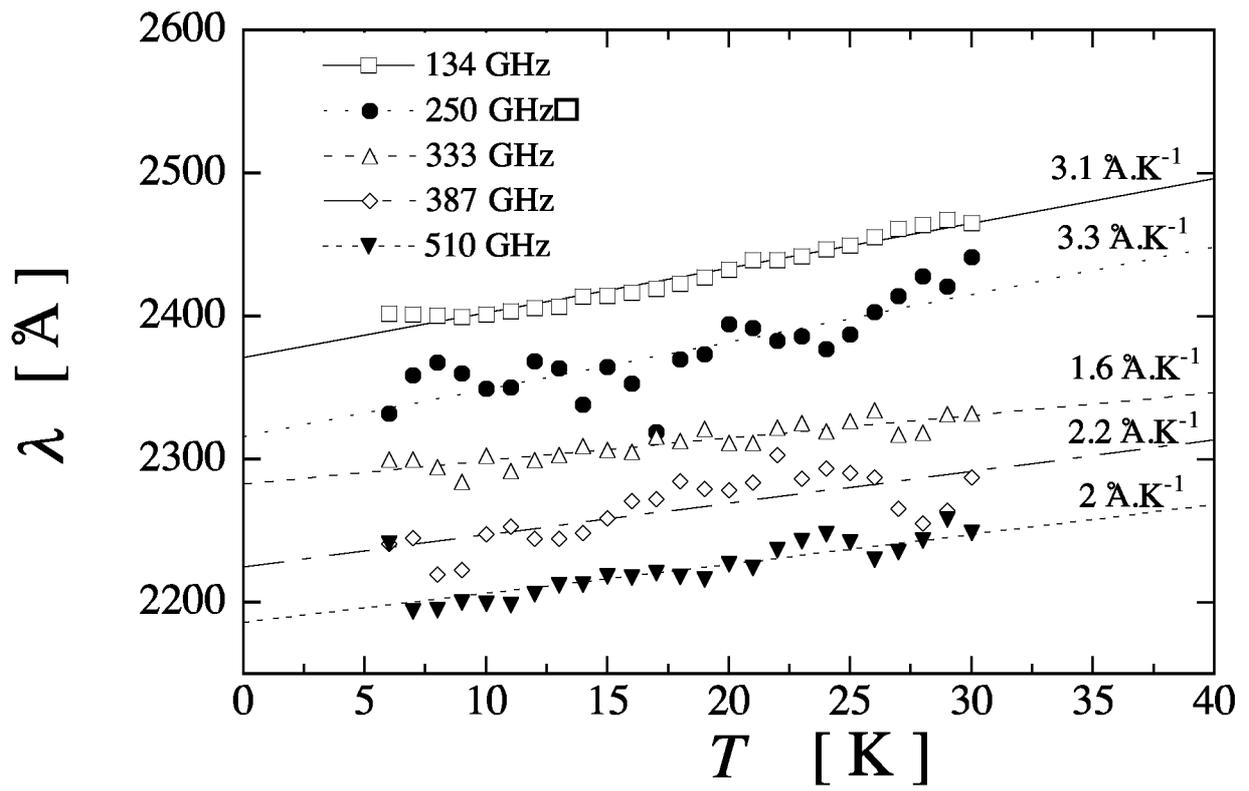
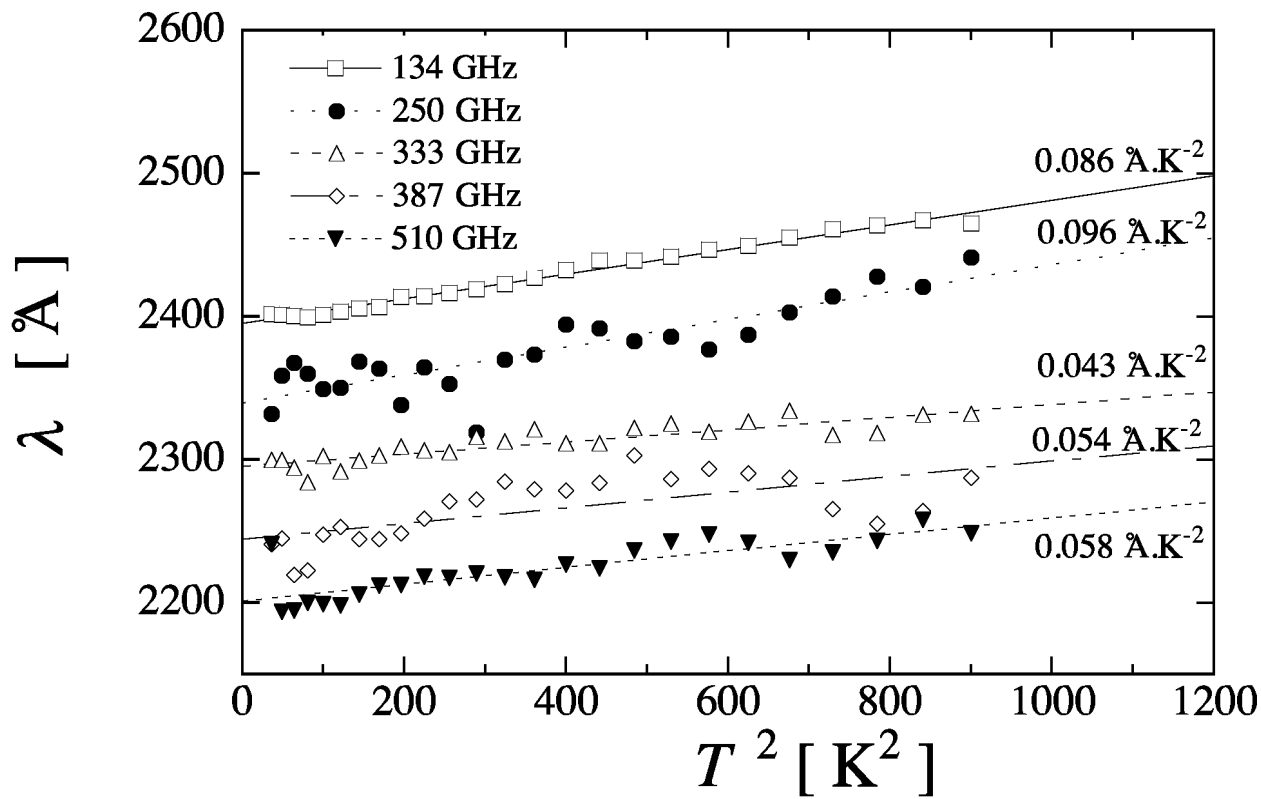
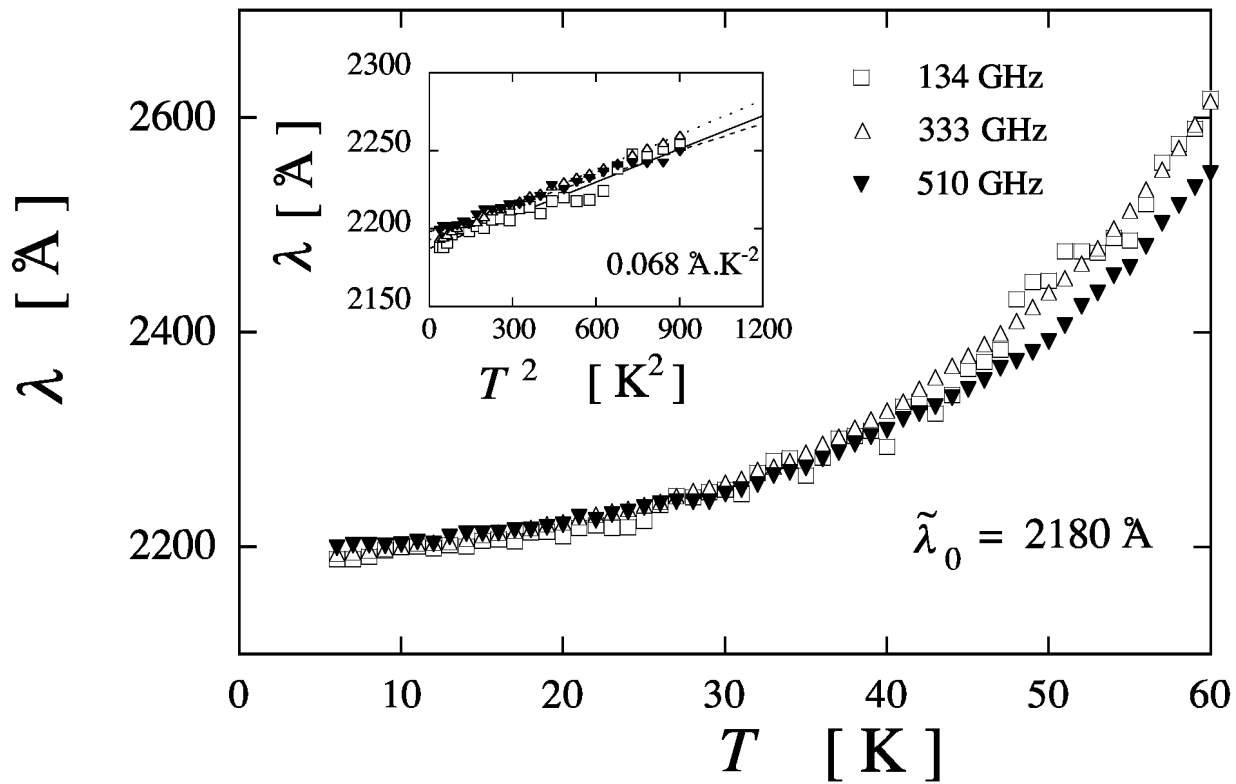
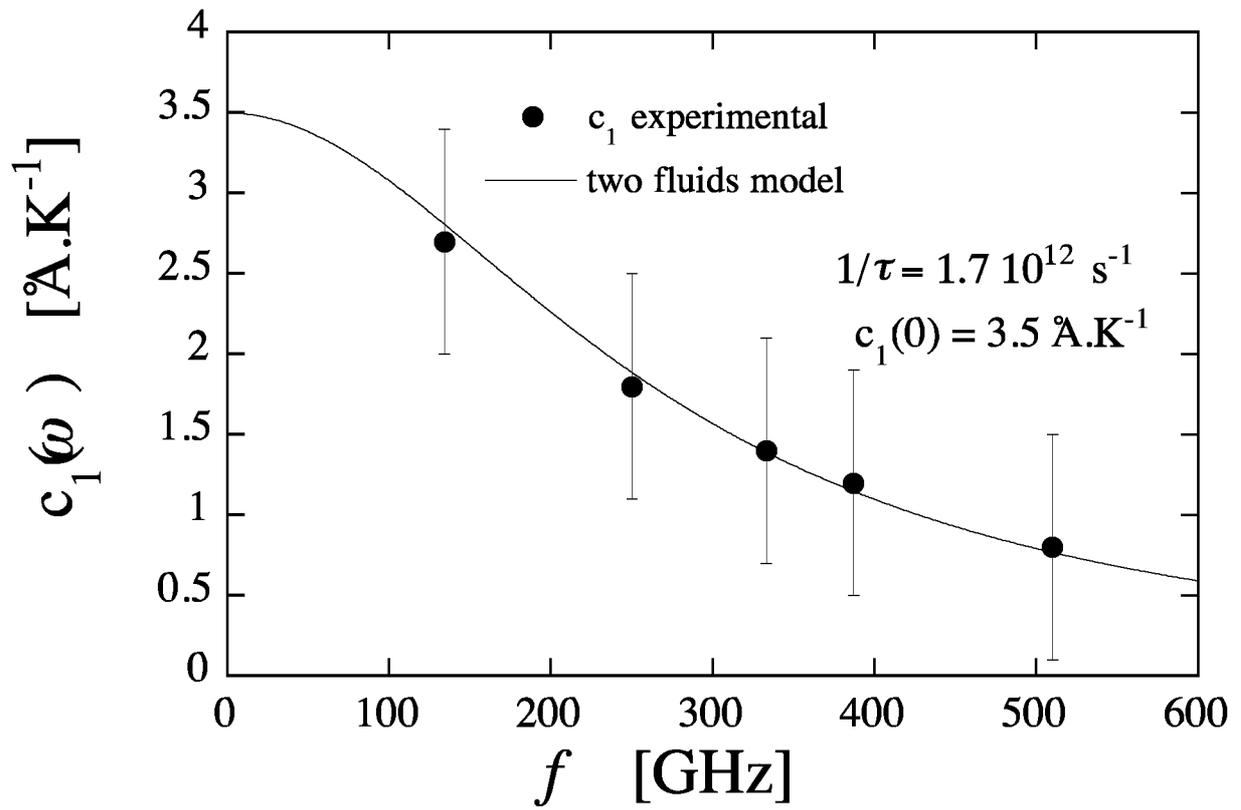
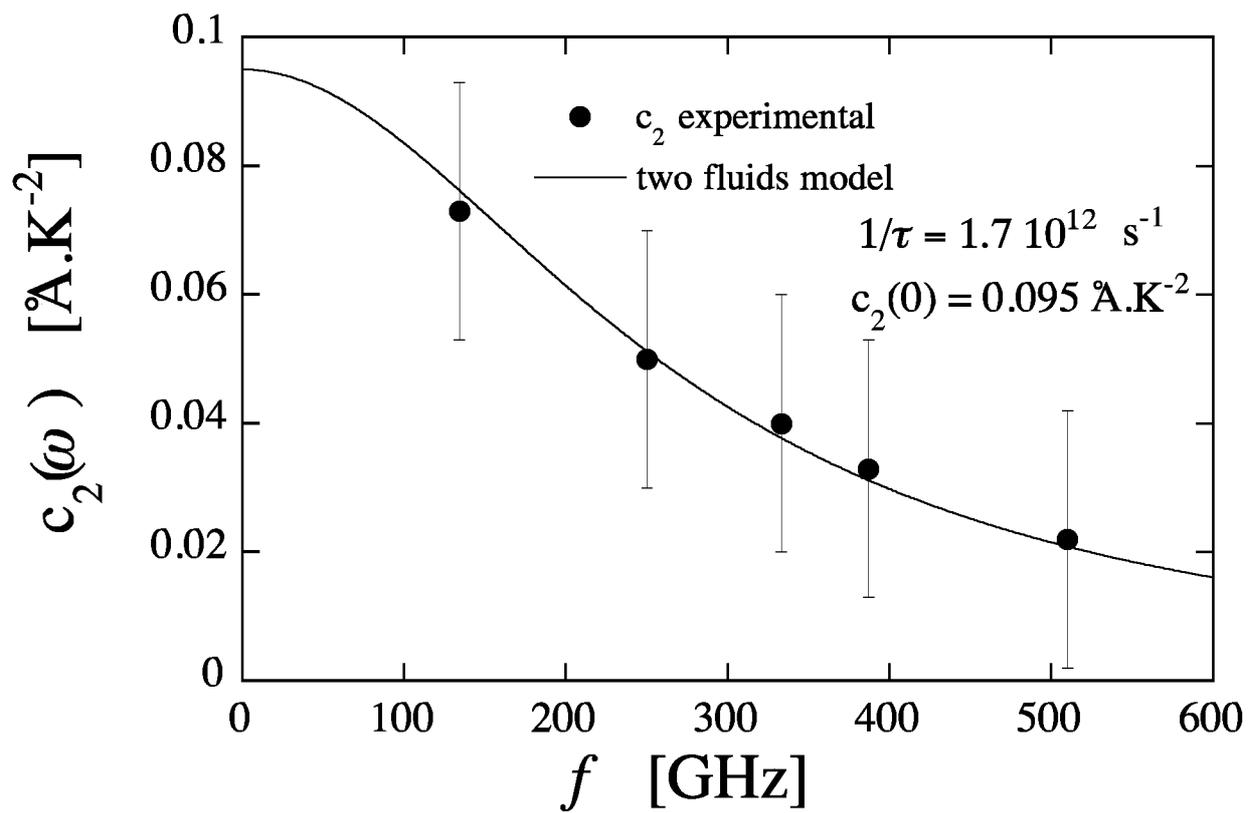
\end{document}